\documentclass{article}
\usepackage{epsfig}
\newcommand{\bfr}{\begin{flushright}}
\newcommand{\efr}{\end{flushright}}
 
\begin{document}
\title{Quantum aspects of self-interacting fields around cosmic strings
}
\author{Kiyoshi Shiraishi\\
Akita Junior College, Shimokitade-Sakura,\\
Akita-shi, Akita 010,
Japan\\
and\\
Satoru Hirenzaki\\
The Institute of Physical and Chemical Research, Wakoh-shi,\\
 Saitama
351-01, Japan}
\date{Class. Quantum Grav. \textbf{9} (1992) pp.~2277--2286}
\maketitle
\begin{abstract}
We study the quantum effect of self-interacting fields in the classical
background of conical space, i.e. around, a cosmic string with
infinitesimal width. The renormalized value of $\langle\phi^2\rangle$ and
energy-momentum tensor in the presence of cosmic strings are calculated
in the self-interacting scalar field theory. The amount of condensation
is also estimated in the case of the Dirac Lagrangian with the
four-fermion interaction. The physical implications of the above
analyses are discussed.
\end{abstract}

\section{Introduction}
The recent enthusiasm for cosmic strings \cite{1} comes from the
possibility that loops of strings provide the seeds for galaxies. The
cosmic strings which have large energy density give seeds for galaxies.
However, the dimensionless combination $G\mu$,%
\footnote{In this paper, we choose the natural unit system in which $c=
h/2\pi=1$.} where $G$ is the Newton
constant and $\mu$ stands for the energy density of the string per unit
length, must not be too large to avoid conflict with astronomical
observation. The most stringent limit is derived from observation of the
microwave background \cite{2}. A network of cosmic strings moving at a
relativistic speed may generate a characteristic pattern of anisotropy
in the temperature of the radiation. The upper limit of anisotropy
places constraints on $G\mu$. According to the result, it is concluded
that `massive' strings which have $G\mu$ much larger than $5\times
10^{-6}$ could not exist in the early universe.

Even in the static case, it is known that the gravitational field
around an infinitely stretching straight string has a peculiar property.
An idealized static string, which has an infinitesimal thickness, cannot
create a Newtonian gravitational potential around it \cite{1,3}. Indeed,
beside the singularity on the string, the spacetime around the string
has vanishing curvature everywhere. The global structure of the space
is, however, non-trivial as many authors advocated.

Suppose the idealized cosmic string lies on the $z$-axis. The metric of
the cylindrically symmetric space can be written \cite{3}
\begin{equation}
ds^2=-dt^2+dr^2+(r^2/\nu^2) d\theta^2+dz^2\,,	
\label{1.1}
\end{equation}
where $\nu^{-1}=1-4G\mu<1$ and $r$ and $\theta$ are the polar
coordinates of the $xy$ plane. If we use a new coordinate
\begin{equation}
\tilde{\theta}\equiv\theta/\nu\,,
\label{1.2}
\end{equation}
we find that the metric reduces to the flat Minkowski spacetime except
for the deficit in the azimuthal angle because $\tilde{\theta}$ takes the
value from zero to $2\pi/\nu$. In this interpretation, it can be
said that the space has a conical singularity at the location of the
idealized cosmic string.

Several authors have studied the quantum field theory around the
conical space \cite{4,5,6,7,8,9}. In many cases, quantum effects are
solved exactly owing to the local flatness of the spacetime.
Furthermore, the quantum aspects around the string attract much
attention as an example of illustrating the importance of global
topology in field theory.

From the physical and astrophysical viewpoint, we must carefully take
the quantum effect around string, because it may be a menace to the
existence of the string itself. The gravitational effect is independent
of the detail of the internal structure of cosmic string, i.e., the
symmetry breaking which produces the topological defects. Therefore, for
example, if the energy density induced from the quantum polarization
becomes enormous, catastrophic increase of the energy density of the
string may occur and at least static straight string-like objects no
longer exist.

According to several authors, so far no terrible effect can be found
due to the free quantum fields around the string. In this paper we
analyse the quantum effects of interacting field theories around the
string. The interaction of fields may enhance or suppress the amount of
the quantum-effect of free fields. We will clarify this point and
discuss its implications.

Although we treat an idealized string in the present paper, an `actual'
string is expected to be made of self-interacting Higgs-like field. Thus
symmetry breaking near the string is an important subject to study. A
study along this line has been made by Russell and Toms \cite{9}. They
did not evaluate physical quantities such as energy density in free space
around the string. We follow the essence of their method in treating
interaction consistently, to calculate the renormalized value of
$\langle\phi^2\rangle$ and the energy-momentum tensor.

In section 2 we calculate the renormalized value of
$\langle\phi^2\rangle$ and the energy-momentum tensor for a massive
scalar. The derivation includes not only a review of the technique to
get the analytic form for the Green function but also detailed
estimations of $\langle\phi^2\rangle$ and the energy-momentum tensor
which have not been exhibited explicitly.

In section 3 we consider the self-interacting massless scalar field.
The effect of interaction is treated self-consistently. The renormalized
$\langle\phi^2\rangle$ and the energy-momentum tensor are calculated.

In section 4 we estimate the amount of condensation of fermions in a
self-interacting model such as the Nambu-Jona-Lasinio model \cite{10,11}.

Section 5 is devoted to a summary and discussion.

\section{The calculation of vacuum fluctuations in massive scalar 
field theory in the conical space}
In this section we demonstrate the calculation of quantum fluctuations
of a massive free scalar field. We first determine the two-point
function of the scalar field, because the renormalized value of
$\langle\phi^2\rangle$ and the energy-momentum tensor due to vacuum
polarization can be derived from the two-point function and a
regularization procedure.

Several authors have developed methods of computing the two-point
functions \cite{4,5,6,7,8,9}. We follow the method given by Smith
\cite{8}, for it seems very simple and applicable.

In his method, we must first solve the normal mode of the wavefunction
in the background of conical space described as (\ref{1.1}). If we write
the mode function as
\begin{equation}
f(r, \theta, z, t)=R(r) e^{im\theta} e^{ikz} e^{-i\omega t}\,,	
\label{2.1}
\end{equation}
we find that the function $R(r)$ obeys the Bessel equation:
\begin{equation}
\left(\frac{1}{r}\frac{d}{dr}r\frac{d}{dr}+(\omega^2-k^2-M^2)-
\frac{\nu^2m^2}{r^2}
\right)R(r)=0\,,
\label{2.2}
\end{equation}
where $M$ is the mass of the scalar. Here we follow the notation of Smith
\cite{8}, except for $\nu$ which stands for his $p$ and the definition of
$O$. He managed to sum up the bilinear of the mode functions on their
`labels'; especially for the radial function the summation is forced to
be an integration. By the same prescription as his, we obtain
\begin{eqnarray}
& &G_\nu(x, x')\equiv\langle 0|\phi(x)\phi(x')|0\rangle\nonumber \\
& &=\frac{\nu}{4\pi^2}
\sum_{m=-\infty}^\infty e^{im(\theta-\theta')}\int_0^\infty	dK\,
K\,J_{\nu|m|}(Kr)J_{\nu|m|}(Kr')K_0[(K^2+M^2)^{1/2}\zeta]\,, 
\label{2.3}
\end{eqnarray}
where $J_p(x)$ is a Bessel function, $K_p(x)$ is a modified Bessel
function of the second kind and $\zeta^2=(z-z')^2-(t-t')^2$. If $M=0$ is
substituted, then it recovers the result of Smith for a massless scalar
\cite{8}.

We can rewrite $K_0[(K^2+M^2)^{1/2}\zeta]$ in an integral representation
\cite{12}: 
\begin{equation}
K_0[(K^2+M^2)^{1/2}\zeta]=\frac{1}{2}\int_0^\infty \frac{dt}{t} \exp
\left(-t-\frac{(K^2+M^2)\zeta^2}{4t}\right)\,.
\label{2.4}
\end{equation}
Using this representation, the integration with respect to $K$ can be
done. Then we obtain \cite{12}
\begin{equation}
G_\nu=\frac{\nu}{4\pi^2\zeta^2}\sum_{m=-\infty}^\infty
e^{im(\theta-\theta')}\int_0^\infty
dt\,I_{\nu|m|}\left(\frac{2rr'}{\zeta^2}t\right)
\exp
\left(-\frac{r^2+{r'}^2+\zeta^2}{\zeta^2}t-\frac{M^2\zeta^2}{4t}\right)
\,,
\label{2.5}
\end{equation}
where $I_\mu(x)$ is a modified Bessel function of the first kind.

To compare this with the result of Linet \cite{5}, we now use the
integral form of $I_\mu(z)$ \cite{12}:
\begin{equation}
I_\mu(z)=\frac{1}{\pi}\int_0^\pi
e^{z\cos\theta}\cos\mu\theta\,d\theta-\frac{\sin\mu\pi}{\pi}
\int_0^\infty e^{-z\cosh x-\mu x}dx\,.
\label{2.6}
\end{equation}
With the aid of the following identity
\begin{equation}
\frac{1}{2\pi}\sum_{m=-\infty}^\infty
e^{im\psi}\cos\nu|m|\theta=\frac{1}{2\nu}
\left[\delta\left(\theta-\frac{\psi}{\nu}\right)+
\delta\left(\theta+\frac{\psi}{\nu}\right)\right]\,,
\label{2.7}
\end{equation}
we can carry out the summation on $m$ and we get:
\begin{eqnarray}
&
&G_\nu=\frac{M}{4\pi^2\{r^2+{r'}^2-2rr'\cos(\varphi/\nu)
+\zeta^2\}^{1/2}}\nonumber \\
& &\times K_1(M\{r^2+{r'}^2-2rr'\cos(\varphi/\nu)
+\zeta^2\}^{1/2})\nonumber \\
&
&+\frac{M\nu}{8\pi^3}\int_0^\infty\frac{K_1(M\{r^2+{r'}^2+2rr'\cosh
t+\zeta^2\}^{1/2})}{(r^2+{r'}^2+2rr'\cosh
t+\zeta^2)^{1/2}}\nonumber \\
& &\times\left(\frac{\sin(\varphi-\nu\pi)}{\cosh\nu
t-\cos(\varphi-\nu\pi)}-\frac{\sin(\varphi+\nu\pi)}{\cosh\nu
t-\cos(\varphi+\nu\pi)}
\right)dt\,,
\label{2.8}
\end{eqnarray}
where $\varphi\equiv\theta-\theta'$.

This result coincides with that computed by Linet \cite{5}. The last
form may be accomplished by any other method, and even in an easier way.

For scalar bosons, it is well known that the vacuum expectation value
of the energy-momentum tensor and $\langle\phi^2\rangle$ can be
calculated from the Green function. We can obtain them for the massive
scalar case using the Green function (\ref{2.8}). 

$\langle\phi^2\rangle$ is the coincidence limit of the renormalized
Green function:
\begin{equation}
\langle\phi^2\rangle=\lim_{x'\rightarrow x} G_{\nu~ren}(x', x)\,,
\label{2.9}
\end{equation}
where $G_{\nu~ren}=G_\nu-G_1$.

Because of the identical nature of divergences in the coincidence limit
in both a cosmic string and the Minkowski background, the subtraction of
the same quantity in Minkowski space assures a finite and meaningful
renormalization scheme.

Similarly the energy-momentum tensor is given by
\begin{equation}
\langle T_\lambda^\mu\rangle=
\lim_{x'\rightarrow x} [T_\lambda^\mu(\nu; x', x)-
T_\lambda^\mu(1; x', x)]\,,
\label{2.10a}
\end{equation}
where
\begin{eqnarray}
& &T_\lambda^\mu(\nu; x', x)\nonumber \\
& &=\left\{
\left[(1-2\xi)\nabla^\mu\nabla_{\lambda'}-\left(\frac{1}{2}-2\xi\right)\delta_\lambda^\mu
\nabla^\rho\nabla_{\rho'}-2\xi\nabla^\mu\nabla_{\lambda}\right.\right.\nonumber
\\ & &\qquad\qquad\qquad\qquad\qquad\left.\left.+M^2
\left(2\xi-\frac{1}{2}\right)\delta_\lambda^\mu\right]G_\nu(x,
x')\right\}
\label{2.10b}
\end{eqnarray}
and $\nabla_\mu~(\nabla_{\mu'})$ denotes the covariant derivative with
respect to the coordinates $x~(x')$. $\xi$ is the coupling to the scalar
curvature. Here we take the conformal coupling, $\xi=1/6$, in the
calculation of the energy-momentum tensor, for simplicity.

The exact expression of $\langle\phi^2\rangle$ for a massive scalar
field is very lengthy, but the estimation in the limiting cases can be
simply carried out.

For $Mr\ll 1$, we obtain
\begin{eqnarray}
& &\langle\phi^2\rangle=\frac{\nu^2-1}{48\pi^2r^2}\,,
\label{2.11a}\\
& &\langle T_\lambda^\mu\rangle=\frac{\nu^4-1}{1440\pi^2r^4}
\mbox{diag.}(1, 1, -3, 1)\,.
\label{2.11b}
\end{eqnarray}
These are the same results as a massless scalar case.

For $Mr\gg 1$, we get
\begin{eqnarray}
& &\langle\phi^2\rangle=\frac{\nu\tan(\pi(\nu-1)/2)}{8\pi^2r^2}
e^{-2Mr}\,,
\label{2.12a}\\
& &\langle T_\lambda^\mu\rangle=\frac{\nu\tan(\pi(\nu-1)/2) M^2
}{24\pi^2r^2}
\mbox{diag.}(-1, 0, -1, -1)\,,
\label{2.12b}
\end{eqnarray}
for $\xi=1/6$ at the leading order in the expansion with respect to the
powers of $(Mr)^{-1}$. We find that exponential damping appears in these
quantities, which is coincident with a naive expectation.

Now the warming-up is over; we will try to consider the inclusion of
self-interaction of bosons in the next section.

\section{Inclusion of self-interaction of scalars}
In this section we consider a massless scalar field theory with a
quartic self-interaction. The Lagrangian density for the theory is
\begin{equation}
L=\frac{1}{2}\nabla_\mu\phi\nabla^\mu\phi+\frac{1}{2}\xi
R\phi^2+\frac{\lambda}{4!}\phi^4\,,
\label{3.1}
\end{equation}
where $\lambda$ is a dimensionless coupling.

One of the reasons for our consideration of the massless case is just
for simplicity. Another reason is as follows; as we have seen in the
last section, the quantum effects are expected to fade out exponentially
with distance from the string. Thus only in the region $Mr\ll 1$ may the
interaction play an important role in the calculation of the physical
quantities.

The interacting field around the cosmic string has been considered by
Russell and Toms \cite{9}. They have taken special care of the boundary
conditions and the instability of each mode of the wavefunction.

In this paper, our aim is to calculate $\langle\phi^2\rangle$ and the
energy-momentum tensor explicitly in the presence of a quartic
interaction without any boundary condition at a finite distance. To this
end, we treat the effect of interaction with an iterative method.

We expect that the vacuum value of $\langle\phi^2\rangle$ is expressed
as
\begin{equation}
\langle\phi^2\rangle=\frac{\alpha}{r^2}\,,
\label{3.2}
\end{equation}
where $\alpha$ is a dimensionless constant, because there is no
dimensional coupling. Now the problem is to obtain the value of
$\alpha$. We try to determine $\langle\phi^2\rangle$ in the following
self-consistent way.

The equation of motion according to the Lagrangian (\ref{3.1}) is
\begin{equation}
-\nabla_\mu\nabla^\mu\phi+\xi
R\phi+\frac{\lambda}{3!}\phi^3=0\,.
\label{3.3}
\end{equation}
The coupling to the curvature is irrelevant to the Green function at
this time, since the spacetime around the cosmic string is locally flat
everywhere except for just on the string.

Here we apply a `Hartree-like' approximation to the equation (\ref{3.3});
we substitute
\begin{equation}
\phi^3\approx 3\langle\phi^2\rangle\phi\,.
\label{3.4}
\end{equation}

Then we can construct the Green function by following the same method
in the previous section. Now the equation of motion is linearized by
substitution of (\ref{3.4}) and (\ref{3.2}); thus the radial function
$R(r)$, similarly defined by (\ref{2.1}), satisfies
\begin{equation}
\left(\frac{1}{r}\frac{d}{dr}r\frac{d}{dr}+(\omega^2-k^2)-
\frac{\nu^2m^2+\alpha'}{r^2}
\right)R(r)=0\,,
\label{3.5}
\end{equation}
where $\alpha'\equiv\frac{1}{2}\lambda\alpha$.

Consequently, we obtain the integral form for the Green function with
yet undetermined $\alpha'$ in this case:
\begin{eqnarray}
& &G_\nu(x, x';\alpha')=\frac{\nu}{4\pi^2}
\sum_{m=-\infty}^\infty e^{im(\theta-\theta')}\nonumber
\\ & &\cdot\int_0^\infty	dK\,
K\,J_{(\nu^2m^2+\alpha')^{1/2}}(Kr)
J_{(\nu^2m^2+\alpha')^{1/2}}(Kr')K_0[(K^2+M^2)^{1/2}\zeta]\,.
\label{3.6}
\end{eqnarray}

Now our plan to estimate $\langle\phi^2\rangle$ and the energy-momentum
in this case is as follows. The derivation
$\langle\phi^2\rangle(\alpha')$ from the renormalized Green function can
be quite parallel to that explained in section 2. The renormalized Green
function is defined as
$G_{\nu~ren}=G_\nu-G_1$. Then we first solve the self-consistent
equation on
$\langle\phi^2\rangle$
\begin{equation}
\frac{\alpha'}{r^2}=\frac{1}{2}\lambda\langle\phi^2\rangle(\alpha')\,.
\label{3.7}
\end{equation}

Next we substitute the value of a' into the expression for the
energy-momentum tensor. This technique is rather like the large $N$
approximation \cite{13}, although $N=1$ in the present case.

Let us return to the actual calculations. For the massless case, the
integration with respect to the variable $K$ can be carried out first and
yields
\begin{equation}
G_\nu(x, x';\alpha')=\frac{\nu}{8\pi^2rr'\sinh u}
\sum_{m=-\infty}^\infty \exp[im(\theta-\theta')]
\exp[-(\nu^2m^2+\alpha')^{1/2}u]\,.
\label{3.8}
\end{equation}
where $\cosh
u\equiv\{r^2+{r'}^2+\zeta^2\}/2rr'=
\{r^2+{r'}^2+(z-z')^2-(t-t')^2\}/2rr'$.
One can see that this reduces to the well known result \cite{4,5,6,7,8,9}
if we let $\alpha'$ go to zero.

Taking the coincidence limit of the renormalized Green function as
(\ref{2.9}), we can reduce (\ref{3.7}) for self-consistency to
\begin{equation}
\frac{\sqrt{\alpha'}}{\lambda}=\frac{1}{8\pi^3}\sum_{n=1}^\infty
\frac{1}{n}\left[\nu K_1\left(\frac{2\pi\sqrt{\alpha'}n}{\nu}\right)-
K_1\left({2\pi\sqrt{\alpha'}n}\right)\right]\,.
\label{3.9}
\end{equation}

This equation is solved numerically. The result is shown in figure 1,
for $\nu=1+10^0$, $1+10^{-2}$ and $1+10^{-4}$. In this figure, we exhibit
$F_1\equiv48\pi^2r^2\langle\phi^2\rangle$ as the result of the solution
of (\ref{3.9}).

For very small $\lambda$
\begin{equation}
F_1\approx\nu^2-1	\,,
\label{3.10}
\end{equation}
whereas for relatively large $\lambda$, $F_1$ decreases in a moderate
way.

The energy-momentum tensor in the present scheme can be written as%
\footnote{We can take other regularization schemes, such as by
subtracting the quantity in which $\lambda$ is fixed (say $\lambda=0$),
which may include logarithmic dependence on $r$; however, the log
correction can be absorbed into renormalization of $\lambda$ and/or is
irrelevant for the present self-consistent approximation scheme.}
\begin{eqnarray}
& &\langle T_\lambda^\mu(\alpha'(\lambda))\rangle=
\lim_{x'\rightarrow x} [T_\lambda^\mu(\nu; \alpha')-
T_\lambda^\mu(1; \alpha')]\,,
\label{3.11a}\\
& &T_\lambda^\mu(\nu; \alpha')\nonumber \\
& &=\left\{
\left[(1-2\xi)\nabla^\mu\nabla_{\lambda'}-\left(\frac{1}{2}-2\xi\right)\delta_\lambda^\mu
\nabla^\rho\nabla_{\rho'}-2\xi\nabla^\mu\nabla_{\lambda}\right.\right.\nonumber
\\ & &\qquad\qquad\qquad\qquad\qquad\left.\left.+\frac{1}{2}
\left(2\xi-\frac{1}{4}\right)\lambda\langle\phi^2
\rangle\delta_\lambda^\mu\right]G_\nu(x,
x'; \alpha')\right\}\,,
\label{3.11b}
\end{eqnarray}
where a `Gaussian' approximation has been adopted. This treatment is
compatible with the previous Hartree approximation. The value for a' is
to be taken as that we obtained as a solution of (\ref{3.9}).

Some manipulations lead to the following result;
\begin{equation}
\langle T_\lambda^\mu\rangle=\frac{1}{1440\pi^2r^4}F_3
\mbox{diag.}(1, 1, -3, 1)
+\frac{1}{24\pi^2r^4}F_1\mbox{diag.}(2, -1, 3, 2)	\,,
\label{3.12a}
\end{equation}
where
\begin{equation}
F_3=90\sum_{n=1}^\infty
\frac{(\alpha')^{3/2}}{\pi n}\left[\nu
K_3\left(\frac{2\pi\sqrt{\alpha'}n}{\nu}\right)-
K_3\left({2\pi\sqrt{\alpha'}n}\right)\right]\,,
\label{3.12b}
\end{equation}
and
\begin{equation}
F_1=12\sum_{n=1}^\infty
\frac{\sqrt{\alpha'}}{\pi n}\left[\nu
K_1\left(\frac{2\pi\sqrt{\alpha'}n}{\nu}\right)-
K_1\left({2\pi\sqrt{\alpha'}n}\right)\right]\,.
\label{3.12c}
\end{equation}
In these expressions, $\alpha'$ is recognized as the solution of
(\ref{3.9}). Consequently, $F_1$ (\ref{3.12c}) is the same as the
previous definition.

Note that the values of both $F_3$ and $F_1$ for finite $\alpha'$ are
less than those for $\alpha'\rightarrow 0$, that is, the limit of
$\lambda=0$.

In figures 1 and 2, numerical values for the functions $F_1$ and $F_3$
are plotted against $\lambda$, respectively.

To summarize this section, we obtain the numerical value of the
renormalized value for $\langle\phi^2\rangle$ and the components of the
energy-momentum tensor in the interacting scalar model with a quartic
coupling. We find $r^2\langle\phi^2\rangle$ becomes smaller for a finite
$\lambda$. The decrease is, however, a moderate one for each case when
the parameter $\nu$ takes a realistic value such that $\nu-1\approx
10^{-6}$. The absolute values of the components of the energy-momentum
tensor always become smaller when $\lambda\ne 0$ and for any $\nu$. Thus
no catastrophic characteristics are found in the interacting boson
theory, at least in the approximation scheme.

In the next section, we will examine the self-interacting fermionic
fields around the string.

\begin{figure}[ht]
\begin{center}
\includegraphics[width=5cm]{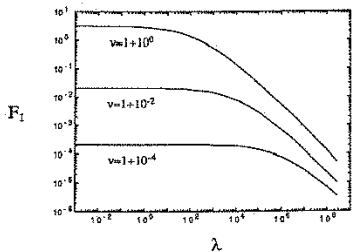}
\caption{$F_1$ plotted against the self-coupling $\lambda$
(see text).}
\label{f1}
\end{center}
\end{figure}

\begin{figure}[ht]
\begin{center}
\includegraphics[width=5cm]{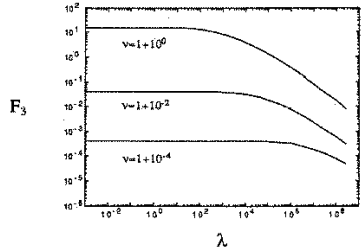}
\caption{$F_3$ plotted against $\lambda$
(see text).}
\label{f2}
\end{center}
\end{figure}

\section{Fermion condensate in four fermion interaction theory near 
the cosmic string}
In this section we investigate an interacting-fermion model. We consider
a four-fermion interacting model in four dimensions. We assume that, as
a physical situation, the thickness of the string is comparable with the
inverse of the scale A at which some fundamental interaction between
fermions is expected. Then we treat the string thickness as
infinitesimal for simplicity, and again use the metric defined in
section 1.

The Lagrangian density under consideration is
\begin{equation}
L=i\bar{\psi^a}\gamma^\mu\nabla_\mu\psi^a+G(\bar{\psi^a}\psi^a)^2	\,,
\label{4.1}
\end{equation}
where the suffix $a$ denotes the `colour' or species, of the fermions,
which is taken as $N$, and the coupling $G$ is of order $\Lambda^{-2}$.
To see whether the dynamical generation of the fermion mass occurs or
not, we examine the following gap equation;
\begin{equation}
M=-\frac{1}{2}G\langle\bar{\psi}\psi\rangle\,, 
\label{4.2}
\end{equation}
where $M$ is the mass of the fermion field, and
$\langle\bar{\psi}\psi\rangle$ is recognized as
\begin{equation}
\langle\bar{\psi}\psi\rangle=N\lim_{x'\rightarrow x} {\rm Tr~} 
S(x, x'; M)	\,,
\label{4.3}
\end{equation}
where $S$ is the propagator (or Green function) of a single fermion field
with mass$=M$.

In the conical background space, we replace the fermion propagator with
$S_\nu$. For a massless case, the fermion Green function in the conical
space is explicitly obtained in \cite{6}. The extension to the massive
case is a tedious but workable task.

In the present paper we will estimate the right-hand side of the gap
equation at the leading order. We note that the large contribution
arises from the region where $Mr\ll 1$, as in the boson case (in
section 2). Further, we assume the spatial variation of the mass of the
fermion is gentle and small.

At the spatial infinity from the origin, the gap equation must be the
same as the flat spacetime;
\begin{equation}
M_\infty=\frac{GN}{8\pi^2}M_\infty
\left(\Lambda^2-M_\infty^2\ln\frac{\Lambda^2}{M_\infty^2}\right)\,,
\label{4.4}
\end{equation}
where $\Lambda$ is the momentum cut-off. On the other hand, at a finite
distance from the string, we have
\begin{equation}
M_r=\frac{GN}{8\pi^2}M_r
\left(\Lambda^2-M_r^2\ln\frac{\Lambda^2}{M_r^2}-\Delta(r, M_r)\right)\,,
\label{4.5}
\end{equation}
where $M_r$ is the mass of the fermion field at the radial distance $r$
from the string and $\Delta(r, M_r)$ is defined as
\begin{equation}
\Delta(r, M_r)=\frac{4\pi^2}{M_r}\lim_{x'\rightarrow x} {\rm Tr~}
 S_{\nu~ren}(x, x'; M_r)\,,
\label{4.6}
\end{equation}
where $S_{\nu~ren}(x, x'; M_r)$ is the renormalized massive fermion
propagator in the conical space. The assumption that the variation of
$M_r$ is slow has been used in (\ref{4.4}).

From (\ref{4.4}) and (\ref{4.5}), the gap equation (\ref{4.5}) has a
non-trivial solution only if
\begin{equation}
\Delta<M_\infty\ln(\Lambda^2/M_\infty^2)	
\label{4.7}
\end{equation}
is satisfied.

We calculate $\Delta$ in the limit $M_rr\ll 1$; otherwise, the vacuum
polarization effect is expected to be suppressed exponentially. The
calculation can simply be done by only use of the massless propagator of
a fermion. The result is
\begin{equation}
\Delta=\frac{4\pi^2}{M_r}\frac{(\nu^2-1)M_r}{24\pi^2r^2}=
\frac{\nu^2-1}{6r^2}\,. 
\label{4.8}
\end{equation}

Then we find that the dynamical mass generation is absent in the
vicinity of the string. The radius of the region is
\begin{equation}
r<r_c=\left(\frac{\nu^2-1}{6M_\infty^2\ln(\Lambda^2/M_\infty^2)}
\right)^{1/2}\,.
\label{4.9}
\end{equation}
In this region, $Mr\ll 1$ is satisfied and the right-hand side of
(\ref{4.9}) is sufficiently large compared to $1/\Lambda$. For instance,
if we take $\nu= 1+10^{-6}$, $\Lambda=10^{15}$GeV and $M=100$GeV, and
then
$\Lambda r_c\simeq 10^9$. Even if the four-fermion interaction is not a
`fundamental' interaction, but induced from the exchange of bosons whose
masses are of order of $\Lambda$, the conclusion does not change because
$\Lambda r_c\gg 1$. Thus our assumption in the approximation scheme is
adequate.

Note that the present phenomenon is only due to the spacetime
structure. Although the thickness of the symmetry restoration is very
small, the length scale along the string may be macroscopic.

\section{Summary and discussion}
In this paper, we have investigated quantum aspects of self-interacting
fields around the idealized string. For the self-interacting scalar
field theory, the vacuum value $\langle\phi^2\rangle$ turns out to
become small when self-coupling exists. The vacuum stress tensor is
reduced in the presence of the self-interaction. Thus no drastic effect
on the very existence of the string-like object is expected.

On the other hand, in the self-interacting fermion theory, the mass
generation of the fermion does not occur very near the string. This
leads to the restoration of the symmetry in some models \cite{11}. In
such models \cite{11} the condensation of gauge fields may take place.
The thickness of the region is very tiny, but the length scale can be
macroscopic along the cosmic string. The effect may have some importance
in the very early universe if it is applied to some complicated
theories. The detailed analyses including numerical results will be
reported in future works.

The back-reaction involving the gravitational field is worth studying.
At the same time, the study of quantum effects of the electromagnetic
fields of superconducting cosmic string \cite{14} is very interesting
when charged matter fields coupled with each other are taken into
account. We hope to report these studies.

\section*{Acknowledgments}
One of the authors (KS) would like to thank J.~Arafune for the essential
discussion on the primitive stage of this work. KS would also like to
acknowledge the financial aid of Iwanami F\=ujukai.

\section*{Note added in proof}
 Since submission of this paper, we have
been informed of the papers \cite{15}, which treat interacting fields
around a cosmic string.

We are very grateful to the referees for helpful comments and
information.


\end{document}